\documentclass[graybox]{svmult}

\usepackage{type1cm}        
\usepackage{makeidx}         
\usepackage{graphicx}        
\usepackage{multicol}        
\usepackage[bottom]{footmisc}

\usepackage{newtxtext}       %
\usepackage{newtxmath}       

\usepackage[flushleft]{threeparttable}
\usepackage{listings}
\usepackage{color}

\usepackage{lscape}

\definecolor{mygreen}{rgb}{0,0.6,0}
\definecolor{mygray}{rgb}{0.5,0.5,0.5}
\definecolor{mymauve}{rgb}{0.58,0,0.82}
\definecolor{myred}{RGB}{179,21,21}
\definecolor{myblue}{RGB}{34,18,181}

\makeindex             

\begin{document}

\title*{Low- and Mixed-Precision Inference Accelerators}
\author{M.J. Molendijk, F.A.M. de Putter, H. Corporaal}
\institute{M.J. Molendijk \at Eindhoven Artificial Intelligence Systems Institute and PARsE lab, Eindhoven University of Technology, PO Box 513, Eindhoven 5600 MB, the Netherlands, \email{m.j.molendijk@tue.nl}
\and F.A.M. de Putter \at Eindhoven Artificial Intelligence Systems Institute and PARsE lab, Eindhoven University of Technology, PO Box 513, Eindhoven 5600 MB, the Netherlands, \email{f.a.m.d.putter@tue.nl}
\and H. Corporaal \at Eindhoven Artificial Intelligence Systems Institute and PARsE lab, Eindhoven University of Technology, PO Box 513, Eindhoven 5600 MB, the Netherlands, \email{h.corporaal@tue.nl}}
%
%
\maketitle

\abstract*{With the surging popularity of edge computing, the need to efficiently perform neural network inference on battery-constrained IoT devices has greatly increased. While algorithmic developments enable neural networks to solve increasingly more complex tasks, the deployment of these networks on edge devices can be problematic due to the stringent energy, latency, and memory requirements. One way to alleviate these requirements is by heavily quantizing the neural network, i.e. lowering the precision of the operands. By taking quantization to the extreme, e.g. by using binary values, new opportunities arise to increase the energy efficiency. Several hardware accelerators exploiting the opportunities of low-precision inference have been created, all aiming at enabling neural network inference at the edge. In this chapter, design choices and their implications on the flexibility and energy efficiency of several accelerators supporting extremely quantized networks are reviewed.}

\abstract{With the surging popularity of edge computing, the need to efficiently perform neural network inference on battery-constrained IoT devices has greatly increased. While algorithmic developments enable neural networks to solve increasingly more complex tasks, the deployment of these networks on edge devices can be problematic due to the stringent energy, latency, and memory requirements. One way to alleviate these requirements is by heavily quantizing the neural network, i.e. lowering the precision of the operands. By taking quantization to the extreme, e.g. by using binary values, new opportunities arise to increase the energy efficiency. Several hardware accelerators exploiting the opportunities of low-precision inference have been created, all aiming at enabling neural network inference at the edge. In this chapter, design choices and their implications on the flexibility and energy efficiency of several accelerators supporting extremely quantized networks are reviewed.}

\section{Introduction}
\label{sec:intro}
Neural Networks can solve increasingly more complex tasks in fields such as Computer Vision (CV) and Natural Language Processing (NLP). While these Neural Networks can perform complex tasks with increasingly higher accuracy, the sheer size of these networks often prevents deployment on edge devices that have limited memory capacity and are subject to severe energy constraints. To overcome the issues preventing the deployment of neural networks onto edge devices, efforts towards reducing the model size and reducing the computational costs have been made. These efforts are most often focused on either the algorithmic side, tailoring the neural network and its properties, or on the hardware side, creating efficient system designs and arithmetic circuitry.

In an effort to reduce the computational cost and model size of neural networks, several approaches are taken. One of these approaches is to tailor the neural network architecture to a specific piece of hardware, this is called hardware-aware neural architecture search (NAS)~\cite{Wu2019FBNET:Search}\cite{Tan2019Mnasnet:Mobile}. Other ways to increase the energy efficiency is by compressing the model size, applying either quantization~\cite{Gholami2021AInference} or pruning~\cite{Blalock2020WhatPruning}. 

In parallel to research on model compression, research has been performed on creating highly-specialized hardware that exploits the opportunities arising from model compression. ASICs that support neural network inference for operand precisions as low as 1-bit exploit the advantages extreme quantization brings: low memory size and bandwidth, and simplified compute logic. In the pursuit of the most energy-efficient hardware design, several design choices regarding memory hierarchy, hardware parallelization of operations, and data-flow are made that impact both the ASIC's efficiency as well as its flexibility.

For instance, many architectures have a fixed-datapath; the movement of the data is fixed at design time which can impose limitations on the layer types, channel dimensions, and kernel dimensions. Furthermore,  these architectures typically have limited programmability and configurability, which restricts the execution schedules that can (efficiently) be run.

In this chapter, a look will be taken at several different approaches used by researchers who specifically designed neural network accelerators for inference with very low-precision operands. The efficiency (and origin thereof) of the architectures will be analyzed and compared to the flexibility that these architectures offer. 

In short, the contributions of this work are:
\begin{itemize}
    \item Overview of state-of-the-art low- and mixed-precision neural network accelerators, in Section~\ref{sec:accel}.
    \item Analysis on the trade-off between the flexibility and energy efficiency of accelerators, in Section~\ref{sec:comp}.
\end{itemize}
The remainder of this chapter is structured as follows, in Section~\ref{sec:bg} background information on neural network architecture and quantization is presented. Thereafter, in Section~\ref{sec:accel}, the low- and mixed-precision accelerators are presented and a comparison is presented in Section~\ref{sec:comp}. Section~\ref{sec:conc} concludes this chapter.

\section{Background: Extreme Quantization \& Network Variety}
\label{sec:bg}
Modern neural network architectures consist of many different layers with millions of parameters and operations. The storage required to store all parameters and features is not in line with the storage capacity typically found on  embedded devices, leading to costly off-chip memory accesses. Next to the memory and bandwidth limitations, computational costs for full-precision (\texttt{float32}) operations require power-hungry compute blocks that quickly overtax the energy requirements of the embedded devices. To reduce both the computational cost as well as the cost of data access and transport, quantization can be applied.

Quantization leads to lower precision parameters and therefore induces information loss. Naturally, when weights and activations can represent fewer distinct values, the representational capabilities of the network decrease. This decrease may create an accuracy loss. In~\cite{Jacob2018QuantizationInference}, Gholami et al. show, however, that quantization down to \texttt{integer8} can be done without significant accuracy loss. But even when quantizing down to \texttt{integer8}, the memory requirements can still overtax the memory capacity typically found in embedded systems. Therefore, research has been done on extreme quantization, i.e. quantization below 8-bit precision. 

In the next subsection, the different building blocks of neural networks are shown. Thereafter, in Section~\ref{sec:bg:binary} and \ref{sec:bg:ternary}, two forms of extreme quantization, namely \texttt{binary} and \texttt{ternary} quantization are discussed. Finally, in Section \ref{sec:bg:mixedprecision} the need for mixed-precision is considered.

\subsection{Neural Network Architecture Design Space}
\label{subsec:nn_arch}

\begin{figure}[t]
    \centering
    \includegraphics[width=0.8\linewidth]{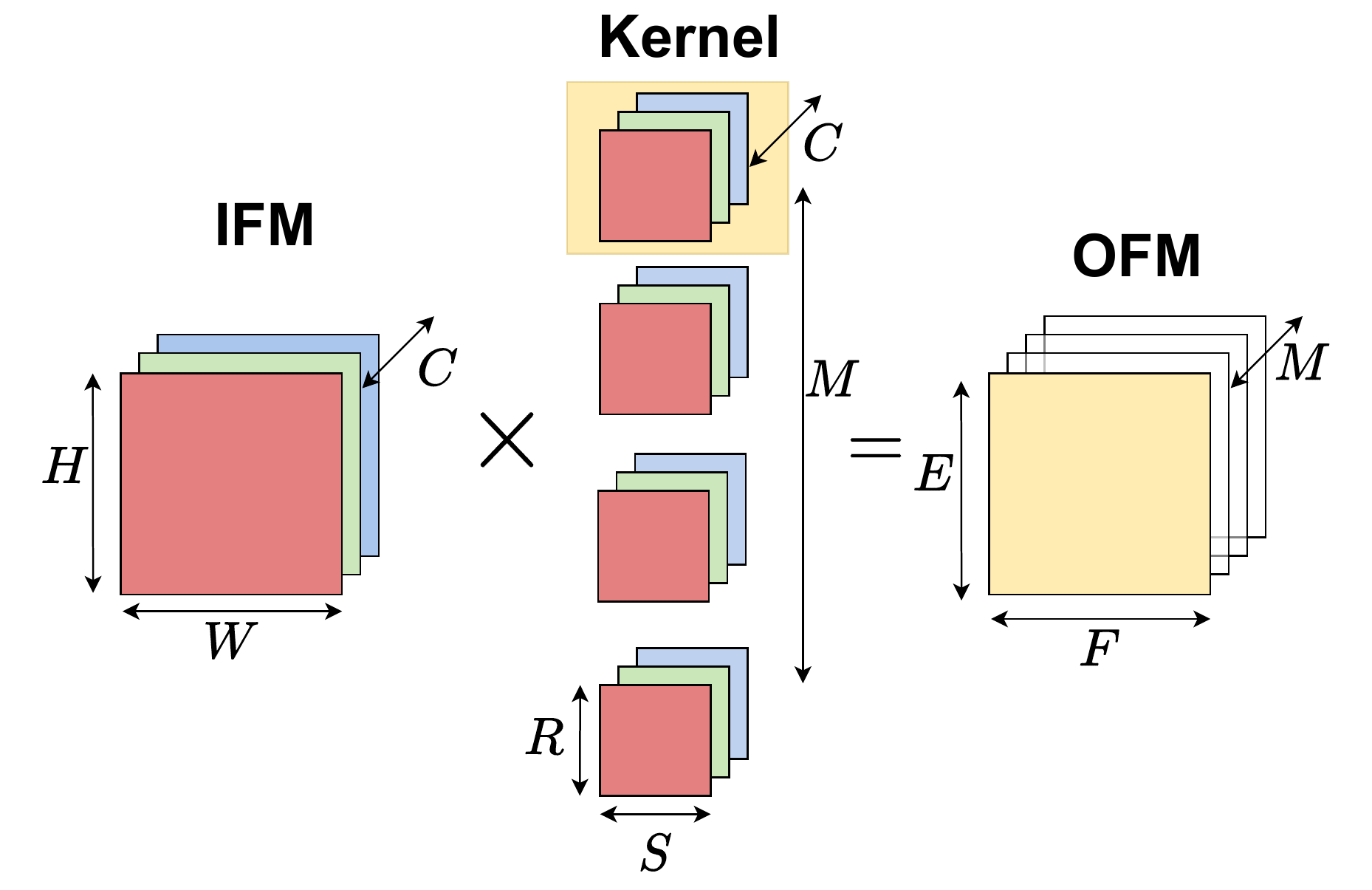}
    \caption{A convolutional layer can vary in different ways. The Input Feature Map (IFM) has height $H$, width $W$ and contains $C$ channels; the Output Feature Map (OFM) has height $E$, width $F$ and $M$ channels; the kernel has height $R$ and width $S$. Between different layers and different networks these parameters vary.}
    \label{fig:conv_params}
\end{figure}
Neural network architectures have a great variety in the type of layers, the size of these layers, and the connectivity between these layers. Furthermore, with mixed-precision architectures, there the precision can also be chosen on a per-layer basis. The neural networks contain different building blocks, the most common ones are listed below.

\begin{itemize}
    \item Convolutional Layer
    \item Fully-connected Layer
    \item Depth-wise Convolutional Layer
    \item Residual Addition
    \item Requantization
    \item Pooling
\end{itemize}

The working horse of most neural network architectures is the convolutional layer. Between different convolutional layers, there can be variety in the kernel size, number of input feature maps, output feature maps, etc. In Fig.~\ref{fig:conv_params}, the different parameters of a convolutional layer are presented. These parameters will later on prove to be an important basis for designing efficient hardware. The goal of Section~\ref{sec:accel} is to show how these network parameters relate to hardware design, hardware efficiency, and hardware flexibility.

\subsection{Binary quantization} \label{sec:bg:binary}
\begin{figure}
    \centering
    \includegraphics[width=0.6\linewidth]{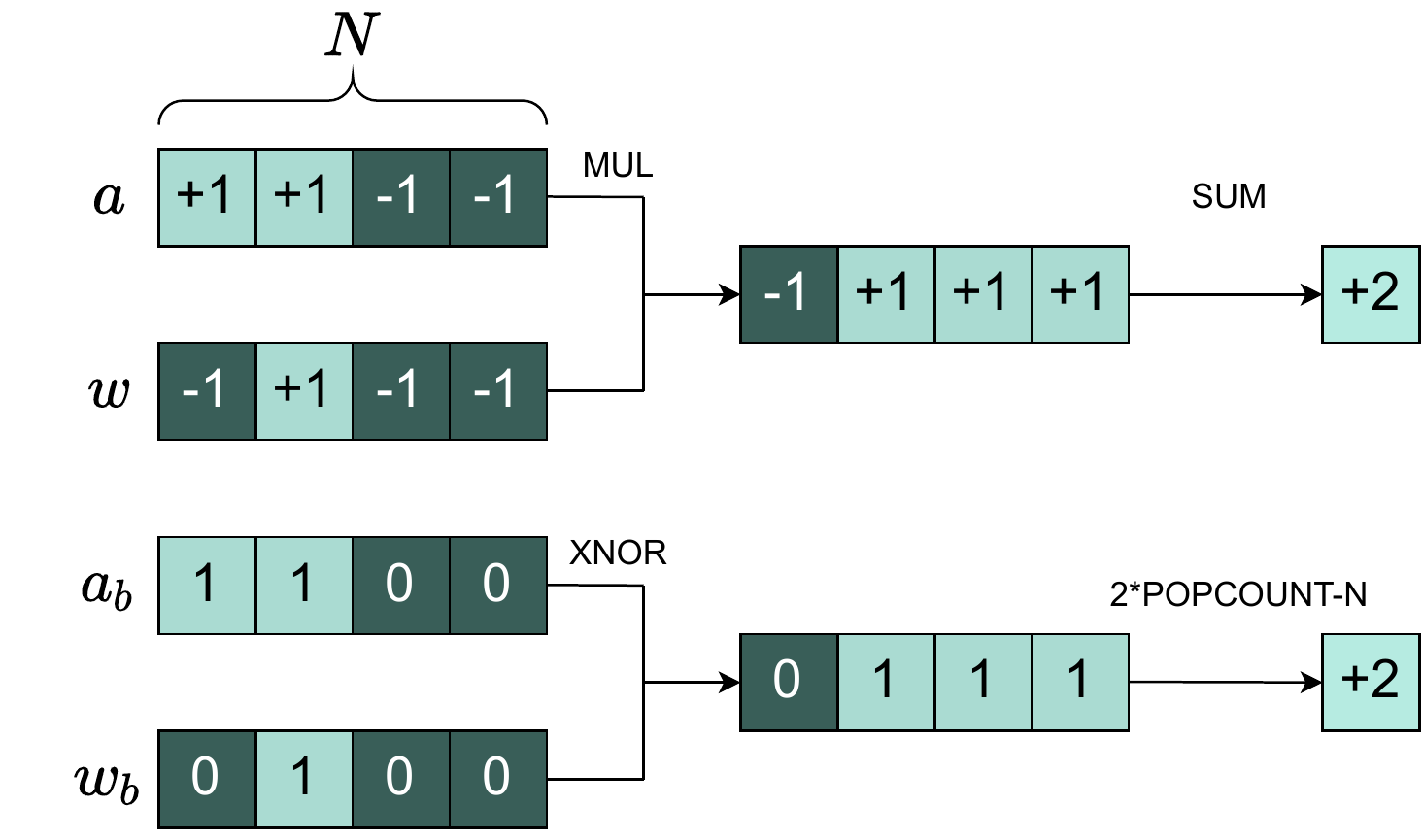}
    \caption{Simplified arithmetic circuitry as a consequence of binary quantization. The top displays the default multiplication while the bottom displays how binary quantization can replace it with \texttt{XNOR} and \texttt{popcount}. N is the number of bits of the input vector.}
    \label{fig:xnorpopc}
\end{figure}

On the extreme end of quantization is binary quantization. Binary quantization restricts both weights and activations to binary. This means that the activations $a \in \{-1, +1\}$ and weights $w \in \{-1, +1\}$. Reduction of the precision of the operands introduces several advantages. First of all, the required storage capacity and bandwidth on the device are drastically reduced, compared to \texttt{float32} by a factor of 32. Furthermore, the Multiply-Accumulate~(\texttt{MAC}) operation, involving expensive multiplication hardware, can be replaced by the much more simple and cheaper \texttt{XNOR} and \texttt{popcount} operations~\cite{Rastegari2016XNOR-net:Networks}. An example of this simplified arithmetic is shown in Fig.~\ref{fig:xnorpopc}.

The output value of the \texttt{popcount} produces a value that needs to be stored with a larger bit-width compared to the binary input value e.g. \texttt{integer16}. Therefore, to feed the outputs into a new layer, the non-linear activation function needs to requantize the values back to the binary bit-width. For this purpose, the \textit{sign} function is used. The quantized operand can be derived from its unquantized form as follows:

\begin{equation}
    X_{quant} = Sign(X) = 
    \begin{cases}
                        +1 & \text{if $X \ge 0$} \\
                        -1 & \text{if $X < 0$} \\
    \end{cases}
\end{equation}

This function is non-differentiable; for training, a Straight-Through Estimator~(STE)~\cite{Bengio2013EstimatingComputation} can be used that passes gradients as is. By employing an STE, gradient descent is possible, and binary neural networks can be trained.

\subsection{Ternary quantization} \label{sec:bg:ternary}
Compared to binary quantization, ternary quantization allows for only one \textit{- albeit very important -} extra value to be represented in the operands, namely zero. Ternary networks therefore have operands $w, a \in \{-1, 0, +1\}$ called \textit{trits}. Next to the increased representational capabilities, the ability to represent zero also solves some issues found in binary networks. First of all, zero padding is not possible in binary networks since it lacks the ability to represent zero, this is most often solved by employing on-off padding. Furthermore, the ability to represent zero introduces the capability to exploit sparsity, i.e. skipping computations when either the weights or activation is zero in a certain case. As will be seen later on, this can have a significant impact on the efficiency of the computational hardware if the network itself is sparse.

\begin{figure}
    \centering
    \includegraphics[width=0.9\linewidth]{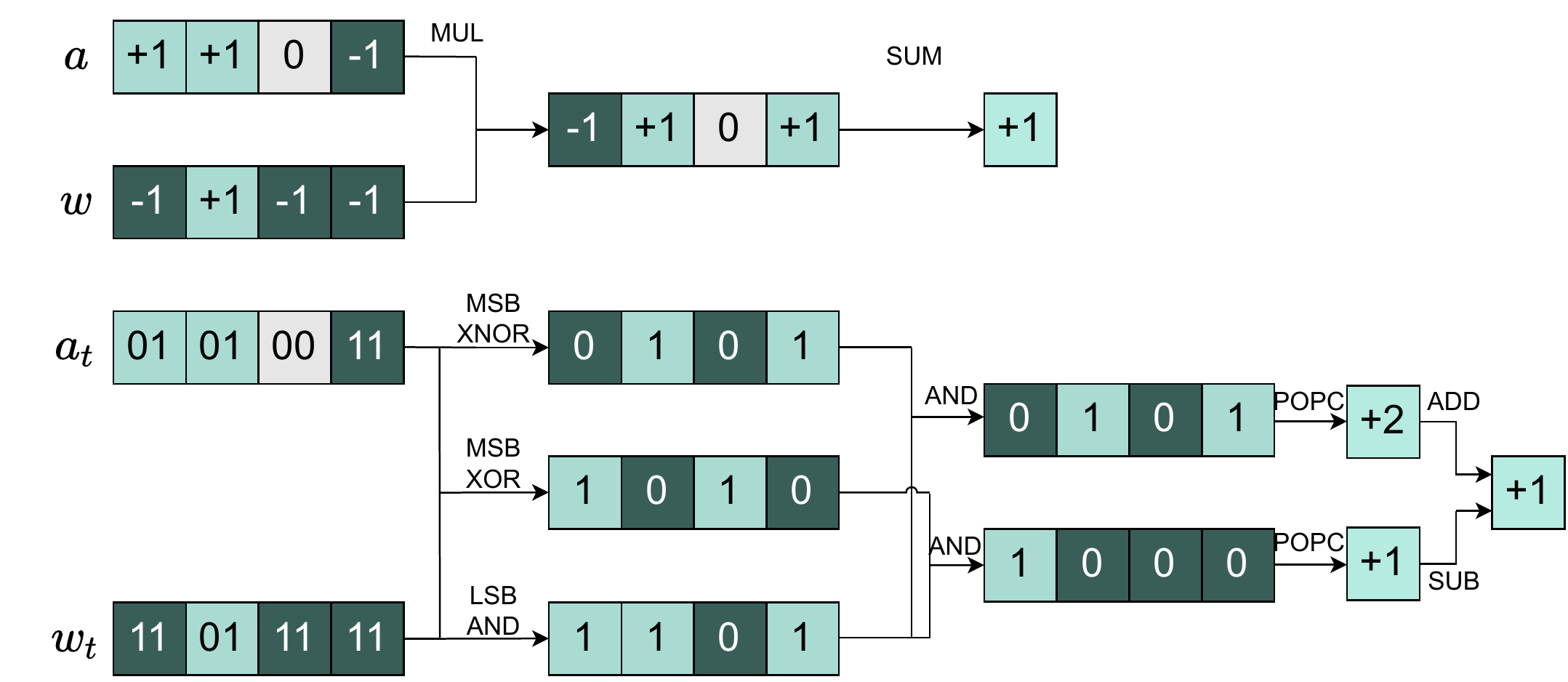}
    \caption{Simplified arithmetic circuitry as a consequence of ternary quantization. The top displays the default multiplication while the bottom displays the ternary simplified variant. Note that the ternary variant needs two popcount modules (one to count +1s and one to count -1s).}
    \label{fig:txnorpopc}
\end{figure}

The arithmetic circuitry required to perform multiply-accumulate (\texttt{MAC}) operations on ternary operands is very similar to that of binary networks. The \texttt{MAC} operation can be replaced by a \texttt{Gated-XNOR}~\cite{Deng2018GXNOR-Net:Framework} (\texttt{XNOR} and \texttt{AND} gates) combined with two \texttt{popcount} modules, one for the +1s and one for the -1s. The arithmetic is shown in Fig.~\ref{fig:txnorpopc}.

Again, as with the binary \texttt{popcount} the final result has a higher bit-width and needs to be requantized before being fed into the next layer. The quantization function typically uses a symmetric threshold value $\Delta$:
\begin{equation}
    X_{quant} = Ternarize(X) = 
    \begin{cases}
                        +1 & \text{if $X > \Delta$} \\
                        0 & \text{if $|X| \leq \Delta$} \\
                        -1 & \text{if $X < -\Delta$} \\
    \end{cases}
\end{equation}

During computation, each trit occupies 2-bits. However, this is a wasteful way to store them since theoretically $log_2(3) = 1.58$ bits are needed for each trit. Muller et al.~\cite{Muller2019EfficientSequences} derived an efficient mapping, compressing 5 trits into 8 bits, yielding a total storage of 1.6~bit per trit, close to the theoretical lower bound.

\subsection{Mixed-Precision} \label{sec:bg:mixedprecision}
Despite all the advantages of extreme quantization, binary and ternary quantization often induce severe accuracy loss, especially on more complex tasks. E.g., there is a large gap in accuracy when comparing \texttt{integer8} quantization to \texttt{binary} and \texttt{ternary}~\cite{Gholami2021AInference}\cite{Bulat2020XNOR-Net++:Networks}. Moreover, the accuracy loss that is induced differs per layer in the network~~\cite{Gluska2020ExploringAnalysis}; i.e. some layers are more resilient to extreme quantization than others. Therefore, a combination of different precisions in a per-layer fashion can give a good balance between accuracy and efficiency.

\begin{figure}[h]
    \centering
    \includegraphics[width=1.0\linewidth]{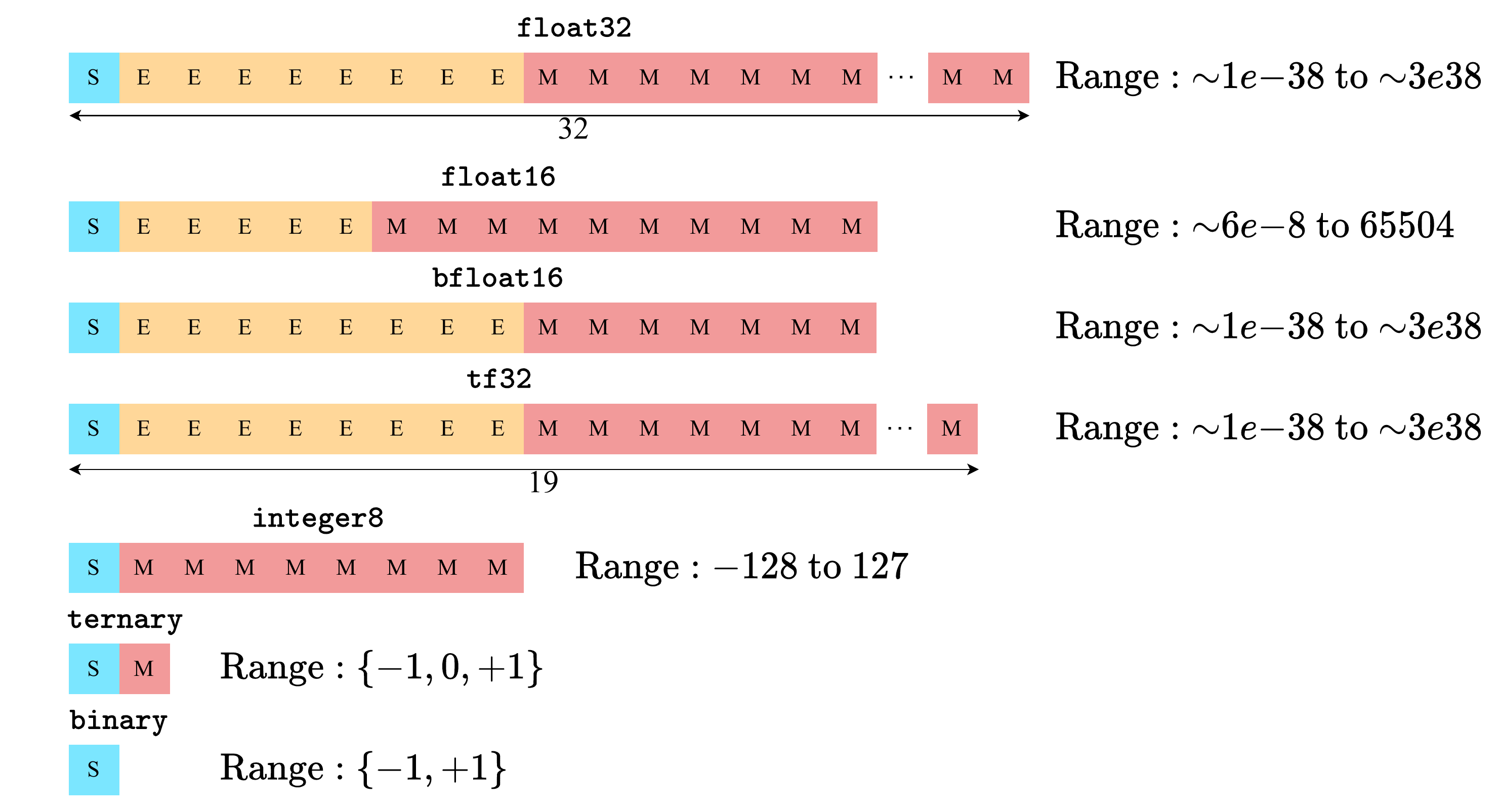}
    \caption{Breakdown of the bit-usage inside data formats commonly used in neural networks. S is sign, E is exponent and M is mantissa. Floating-Point data formats specifically for neural networks prefer higher range over more precision.}
    \label{fig:data_formats}
\end{figure}

An overview of different data precisions typically found in neural network architectures is given in Fig.~\ref{fig:data_formats}. The figure shows the width of different data formats, and how the bits are allocated. Next to the data format, the range is displayed, i.e. the minimum and maximum value that can be attained using that data format. Note that the range for the floating-point number only displays the positive numbers, while it is able to represent negative numbers using the sign bit.

In the past, \texttt{float32} was used as the \textit{de facto} standard for neural networks. Gradually, movements toward smaller data types like \texttt{float16} were made to save on storage and computational cost. Moreover, it was found that the dynamic range of the data types has a larger impact on the accuracy than the relative precision, leading to the creation of \texttt{bfloat16}~\cite{Wang2019BFloat16:TPUs} (Brain Floating Point) and \texttt{tf32}~\cite{Kharya2020TensorFloat-3220x} (TensorFloat32), both trading off relative precision in favor of increased range. Using \texttt{integer8} precision completely gets rid of the expensive floating-point arithmetic, vastly increasing the throughput and energy efficiency, with at the extreme end, \texttt{binary} and \texttt{ternary} quantization.

By nature of floating-point arithmetic units, exponents are added up together while the mantissa bits are multiplied. Therefore, \texttt{bf16}, which has 3 less mantissa bits compared to \texttt{float16} will have a \textit{two times} smaller footprint, while compared to \texttt{float32} it will even have an \textit{eight times} smaller area . This is because the area of the multiplier unit is roughly proportional to the square of the mantissa bits. In Section~\ref{sec:accel}, accelerators that support \texttt{integer8} (which can also be used for fixed-point arithmetic), \texttt{binary} and \texttt{ternary} precisions are discussed.

\section{Accelerators for Low- and Mixed-Precision inference}
\label{sec:accel}

With the aim to get the energy per \texttt{MAC} operation as low as possible, several accelerators specifically designed for low-precision inference have been created. Some of these architectures also support different precisions on the same platform. The accelerators can be split into two groups: fully-digital accelerators and mixed-signal and analog compute-in-memory (CIM) approaches. Although state-of-the-art CIM architectures~\cite{Valavi2019ACompute}\cite{Bankman2019AnCMOS} and mixed-signal implementations~\cite{Ueyoshi2022DIANA:SoC} have the potential to achieve high energy efficiency, they introduce their own challenges. These challenges include longer design time and chip-to-chip variation due to CMOS process variation which makes it more difficult to benchmark the actual performance of such a design, and no programmability, making it more difficult to use the accelerator. The further focus in this chapter will therefore be solely on fully digital implementations.

First, characterization criteria that are important to embedded neural network accelerators will be established, these include key performance indicators, to measure the efficiency (in both area and energy) of the architecture. Furthermore, the basis for the flexibility analysis is laid out, based on the robustness of architectures against different layer types, dimensionality, and precisions. Thereafter, five state-of-the-art digital inference accelerators will be discussed.

\subsection{Characterization Criteria}
\label{subsec:charac}

The accelerators will be characterized both according to their flexibility as well as their energy efficiency. Defining flexibility as a quantitative metric can often be cumbersome, although some recent effort toward bringing structure has been made~\cite{Huang2022HowSystem}. Next to the flexibility aspects, the most important quantitative performance evaluation criteria for neural network inference accelerators will be listed and motivated.

\subsubsection{Flexibility}
\label{sec:flex}
Before the characterization criteria are established, a closer look is taken at the nature of a convolution kernel. A convolution kernel can be described by 6 nested for loops (7 when adding the batch dimension), an exemplary schedule is shown in Listing~\ref{lst:naive_cnn}. It is assumed that the target application is image processing, i.e. inputs are referred to as \textit{pixels}.
\newcommand\Comment{\hfill\normalfont\itshape}
\begin{lstlisting}[language=python, caption={A naive convolutional layer with output-stationary schedule and a stride of 1; \textit{acc} is the temporary accumulated value. The loop iterators are visualized in Fig.~\ref{fig:conv_params}.}, numbers=none, label=lst:naive_cnn, frame=tb]
for h in [0, E]: @:\Comment \textcolor{myred}{Output feature map \textbf{height}} :@
  for w in [0, F]: @:\Comment \textcolor{myred}{Output feature map \textbf{width}} :@
    for m in [0, M]: @:\Comment \textcolor{myred}{\textbf{Ouput} channels} :@
      acc = bias[m]
      for c in [0, C]: @:\Comment \textcolor{myred}{\textbf{Input} channels} :@
        for r in [0, R]: @:\Comment \textcolor{myred}{Kernel \textbf{height}} :@
          for s in [0, S]: @:\Comment \textcolor{myred}{Kernel \textbf{width}} :@
            acc += ifm[h + r][w + s][c] * weights[n][r][s][m]
      ofm[e][f][m] = acc
\end{lstlisting}
In the example, the loops are arranged in a so-called \textit{output stationary} way, i.e. one output pixel is calculated as soon as possible. In other words, all calculations needed for a set of output pixels are performed before moving to the next set of output pixels. This avoids having to store and re-load partially calculated output pixels.

Loop nest optimization (LNO) can be performed to increase data locality. Two important techniques, part of LNO, are loop tiling (also known as loop blocking), where a loop is split up into an inner and outer loop, and loop interchange, where two loops are swapped in hierarchy level. The problem of finding the best combination of the two is called the \textbf{temporal mapping} problem (i.e. finding the best execution schedule). The temporal mapping greatly influences the number of memory accesses needed and therefore indirectly greatly influences the energy efficiency of an accelerator. 

Next to the temporal mapping, the operations performed in the convolutional kernel can also be parallelized in hardware. The problem of finding the optimal parallelization dimensions is called the \textbf{spatial mapping} problem. Using optimal spatial mapping can increase data reuse in hardware and reduce memory traffic. A good example of this is the mapping on a systolic array. It is important to note that the spatial mapping should be carefully chosen, as it imposes constraints on the dimensions being parallelized. 

Hardware parallelization over a dimension is called vectorization. Vectorization over any of the dimensions given in Fig.~\ref{fig:conv_params} will be denoted as the vectorization factor $v_{param}$ where param can be any of the dimensions in Fig.~\ref{fig:conv_params}. For instance, when parallelizing over the $C$ dimension using a vectorization factor of 32, it is denoted as $v_C = 32$. This vectorization factor also implies constraints: any convolutional network layer that does not have an input channel multiple of 32 will not run at 100\% utilization. There will be a trade-off between the vectorization factor and the flexibility with respect to convolutional layers with certain layer dimensions being able to run at full utilization.

Research has been done on structurally exploring the temporal and spatial mapping design space~\cite{Parashar2019Timeloop:Evaluation}~\cite{Wu2019Accelergy:Designs}. Most recently the ZigZag framework~\cite{Mei2021ZigZag:Accelerators} has been published aiming to fully co-design temporal mapping with hardware architecture finding the best spatial and temporal mappings available. 

One other facet of flexibility is \textbf{progammability}. Programmability allows running different, even non-DNN workloads on the accelerators. Furthermore, high-level programmability increases the usability of the device since it allows the workload to be configured while programming it via a high-level language.

\subsubsection{Performance Characteristics}
\label{sec:kpi}
To compare the performance of the several accelerators reviewed, some quantitative metrics that reflect the performance of the accelerator are established. First of all, the most widely promoted metric to compare accelerators is to compare the energy efficiency, defined as the energy per operation (either [pJ/op] or inversely in [TOPS/W]).

Secondly, the memory capacity. Memory capacity plays an important role in the efficiency of the accelerator. Since off-chip memory access energy is much larger than the energy needed to compute, off-chip memory access should be avoided at all costs. More on-chip memory means fewer external memory accesses, benefiting energy consumption. Two different ways to implement on-chip memory are SRAM and Standard-Cell Memory (SCM). While SRAM has a much higher memory density, it is less efficient in terms of energy usage for smaller sizes compared to SCM. Especially when applying voltage-frequency scaling, the SCM can be scaled to a much lower voltage than SRAM. Therefore, SCM tends to be a popular choice to keep down the energy cost of the total system while sacrificing area and storage capacity. Other important metrics are throughput~[GOPS] and area efficiency~[GOPS/mm$^2$].

\subsection{Five Low- and Mixed-Precision accelerators reviewed}
\label{subsec:review}
Five state-of-the-art accelerators will be discussed and compared against each other. These accelerators were chosen because of their support for very low precisions (i.e. \texttt{binary} or \texttt{ternary}). These accelerators are:
\begin{itemize}
    \item \textit{XNOR-Neural Engine}~\cite{Conti2018XNORInference} is a binary neural network accelerator built into a programmable microcontroller unit. A full system-on-a-chip (SoC), implemented in 22nm technology is presented including the accelerator, RISC host processor, and peripherals.
    \item \textit{ChewBaccaNN}~\cite{Andri2020ChewBaccaNN:Accelerator} is an architecture for binary neural network inference that exploits efficient data reuse by co-designing the memory hierarchy with the neural network ran on the architecture. The hard-wired kernel size allows efficient data reuse.
    \item \textit{CUTIE}~\cite{Scherer2020CUTIE:Efficiency} is an accelerator for ternary neural networks. This is a massively parallel architecture, hard-coding all the network parameters into the hardware design. Furthermore, it exploits sparsity opportunities from ternary networks that are not present in binary networks.
    \item \textit{Knag et al.} produced a binary neural network accelerator in 10nm FinFet technology~\cite{Knag2021ACMOS}. The design focuses on utilizing the compute-near-memory paradigm, minimizing the cost of data movement by interleaving memory and computational elements.
    \item \textit{BrainTTA} is a flexible, fully-programmable solution based on a Transport-Triggered Architecture. The architecture has support for mixed-precision and focuses, next to the energy efficiency objective also on flexibility, trying to minimize the concessions made while still pursuing energy efficiency.
\end{itemize}

\subsubsection{XNOR Neural Engine (XNE)}
\begin{figure}[htb]
    \centering
    \includegraphics[width=0.74\linewidth]{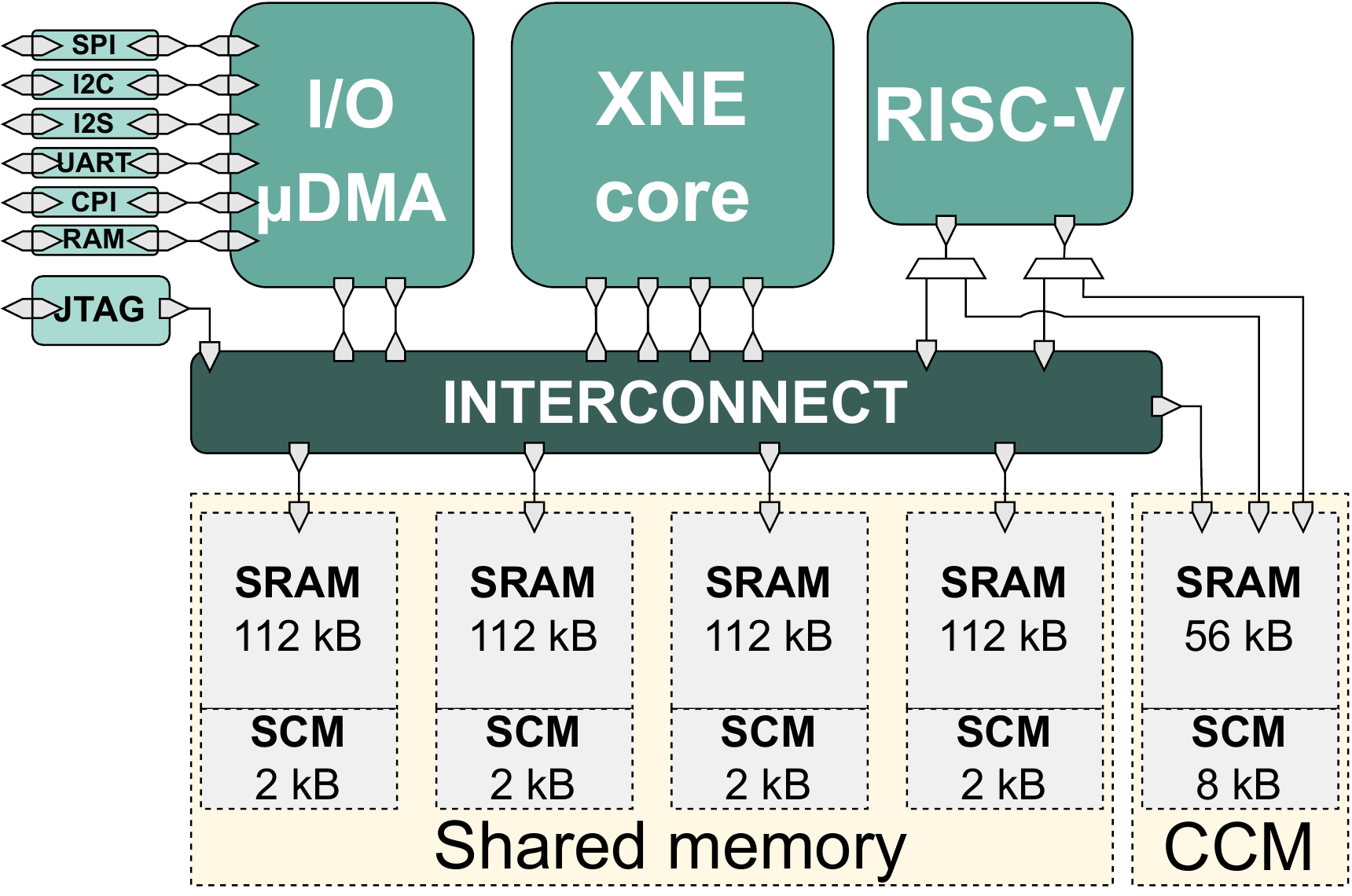}
    \caption{Top-level view of the SoC with XNE inside the MCU. The memory is a hybrid of latch-based SCM and SRAM.}
    \label{fig:xne}
\end{figure}

XNOR-Neural Engine~\cite{Conti2018XNORInference} is a binary accelerator exploiting the arithmetic simplifications introduced by binarizing the weights and activations (see Fig.~\ref{fig:xnorpopc}). Conti et al. present an SoC consisting of an accelerator core (XNE) inside a microcontroller unit (MCU) and peripheries. The accelerator can independently run simple network configurations but requires the programmable MCU to execute more complex layers. The MCU is programmed using some assembly dialect. The full system is shown in Fig.~\ref{fig:xne}. It consists of:

\begin{itemize}
    \item \textit{XNE core}, where the \texttt{binary MAC} operations are performed; this core consists of a \textit{streamer}, to stream feature maps and weights in and out of the architecture, a \textit{controller} consisting of a finite-state machine, the programmable microcode processor and a latch-based register file.
    \item \textit{RISC-V host processor}, used to realize more complex layer behaviors than supported with the XNE core alone.
    \item \textit{Shared Memory}, shared between the $\mu$DMA, RISC-V core, and XNE core. This memory is a hybrid of SRAM and SCM, allowing aggressive voltage scaling when the SRAMs are turned off.
    \item \textit{Core-Coupled Memory (CCM)}, primarily for the RISC-V core, again composed of both SRAM and SCM.
    \item \textit{$\mu$DMA}, which is an autonomous unit able to send and receive data via several communication protocols from and to the shared memory.
\end{itemize}

\begin{figure}[htb]
    \centering
    \includegraphics[width=1.0\linewidth]{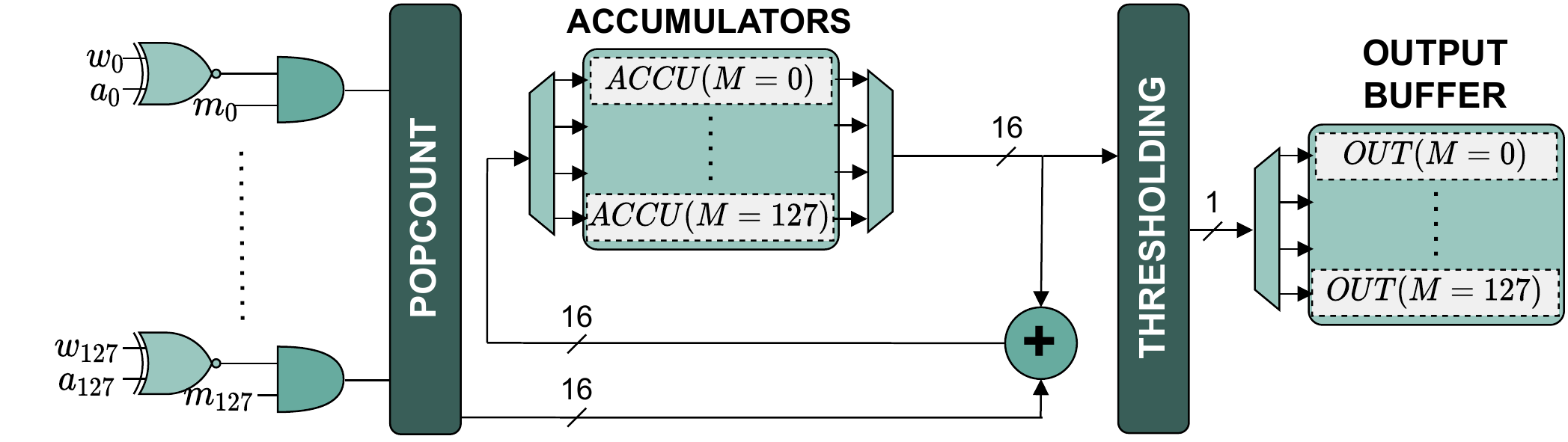}
    \caption{Accelerator core of XNOR Neural Engine with \texttt{TP} = 128. The XNOR operation is performed on the activations $a$ and weights $w$. Whenever the number of input operands is not a multiple of \texttt{TP}, the outputs can be masked by masking bits $m$ to make sure they do not contribute to the popcount output.}
    \label{fig:xneCore}
\end{figure}

The accelerator core, XNE, is shown in Fig.~\ref{fig:xneCore}. The throughput of the design can be chosen at design-time by means of a throughput parameter \texttt{TP}. This throughput parameter can be described as follows: it takes the accelerator \texttt{TP} cycles to calculate \texttt{TP} output pixels. While doing this, the accelerator keeps the same input activations for \texttt{TP} cycles while loading \texttt{TP} weights each cycle (for a total of \texttt{TP} sets of \texttt{TP} weights). Therefore, this \texttt{TP} parameter essentially hard-wires the $C$ and $M$ dimension of the convolution dimensions shown in Fig.~\ref{fig:conv_params} into the design. 

For instance, each accumulator in Fig.~\ref{fig:xneCore} contains the partial result of one output pixel (i.e. the number of accumulators is equal to the output feature map channel vectorization $v_M$). Therefore, all the inputs that are processed while a single accumulator is selected via the mux should contribute to the same output pixel. In this case, the different pixels concurrently offered to the compute core belong to different input channels. Therefore, the choice of \texttt{TP} directly imposes a constraint on the $C$ and $M$ loops in order to run at full efficiency. Furthermore, the output of the \texttt{popcount} operation is directly fed through the binarization function; this means that partial (higher bit-width) results can not be extracted, prohibiting their use for residual layers. For benchmarking the platform a \texttt{TP} factor of 128 was chosen, this means that $v_C = 128$ and $v_M = 128$ for this design-point.

\subsubsection{ChewBaccaNN}

\begin{figure}[htb]
    \centering
    \includegraphics[width=0.70\linewidth]{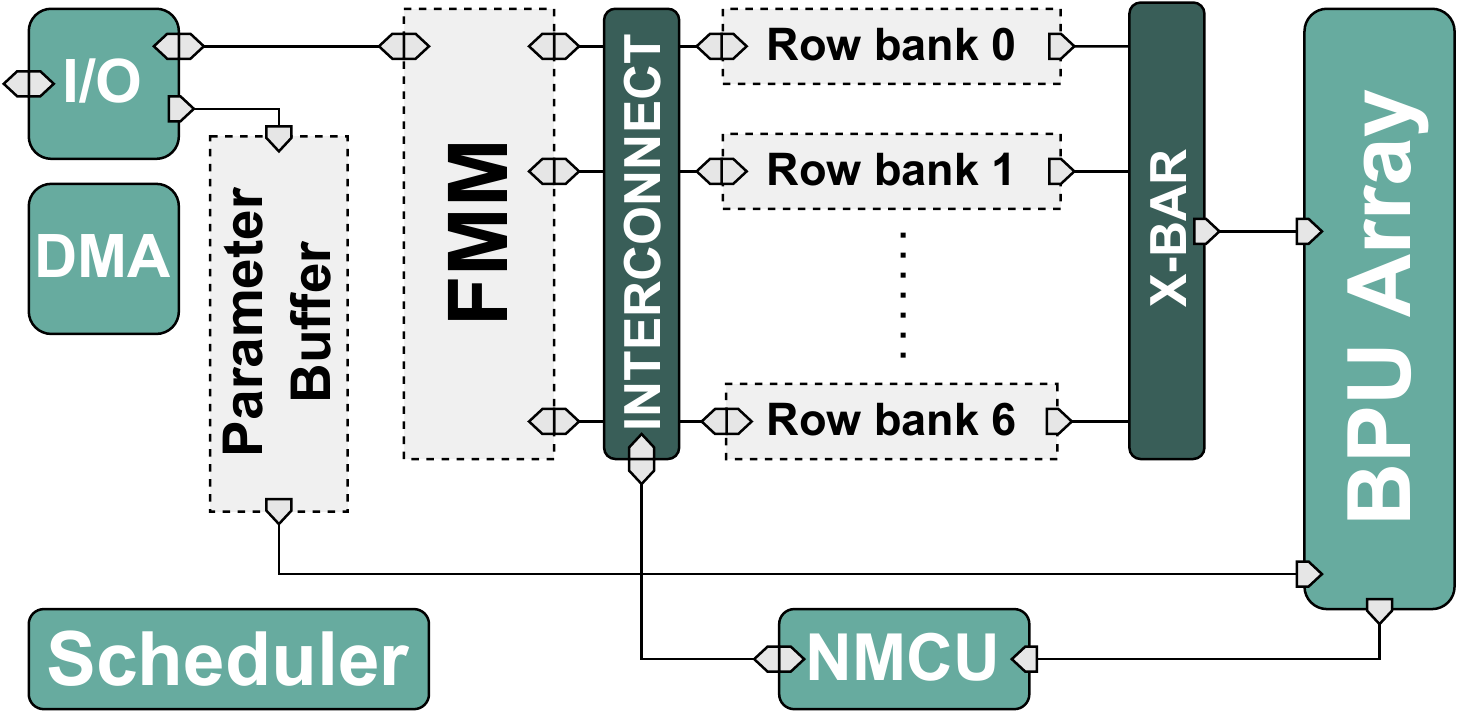}
    \caption{The top-level architectural overview of the ChewBaccaNN accelerator. All the memories are implemented using latch-based standard-cell memory. The control signals are not shown in this overview.}
    \label{fig:chewbaccaNN}
\end{figure}

\begin{figure}[htb]
    \centering
    \includegraphics[width=0.8\linewidth]{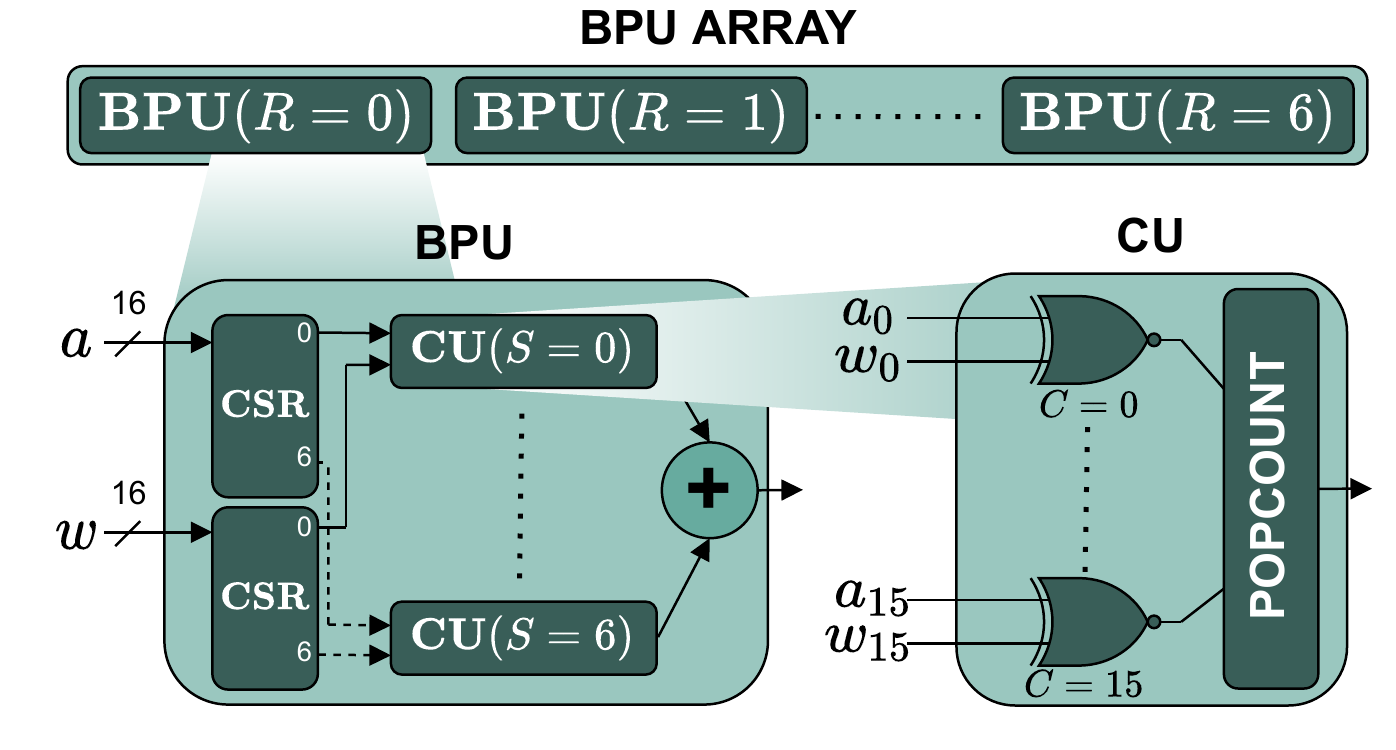}
    \caption{ChewbaccaNN compute core. Hardware parallelization is performed over the kernel height ($R$) in the Basic Processing Unit (BPU) array, over the kernel width ($S$) inside a single BPU and over the input channel dimension ($C$) inside the compute unit (CU). The Controlled Shift Register (CSR) enables data reuse in a sliding window fashion. The architecture contains a total of $16 \times 7 \times 7 = 784$ ($v_C \times v_R \times v_S)$ binary multipliers.}
    \label{fig:chewbaccaNNCore}
\end{figure}

ChewBaccaNN~\cite{Andri2020ChewBaccaNN:Accelerator} is like XNE an accelerator utilizing binary weights and binary activations. Contrary to XNE, this architecture does not implement a full SoC and is therefore purely based on the accelerator core. ChewBaccaNN uses SCM to aggressively scale the voltage down. The design is implemented using GF 22 FDX technology. A top-level view of the architecture is shown in Fig.~\ref{fig:chewbaccaNN}. The components in this architecture are:
\begin{itemize}
    \item \textit{BPU Array} consists of seven Basic Processing Units (BPUs) and forms the computational heart of the accelerator; the BPU is detailed in Fig.~\ref{fig:chewbaccaNNCore} and is discussed in the next paragraph.
    \item \textit{Feature Map Memory (FMM)} holds the input and output feature maps and also has the ability to store partial results (e.g. for residual layers). The FMM is implemented using SCM only. This enables aggressive voltage scaling for the whole chip while sacrificing memory capacity.
    \item \textit{Row Banks} buffer the input feature map rows and kernel rows. \textit{crossbar} (x-bar) is utilized when the sliding convolutional window down. Since each BPU processes one kernel row, the weights can stay inside the BPUs while the input feature map needs to move one row down. This is done by loading one new row and shifting the other rows by one BPU (using the crossbar).
    \item \textit{Scheduler}, used to control the crossbar behavior and make sure that the row banks are timely rotated to the next BPU and the correct weights and IFM pixels are loaded.
    \item \textit{Near Memory Compute Unit (NMCU)}, writes output data from the BPU array to the correct location in the FMM, accumulates residual paths, rebinarizes results, and is used for bit-packing (rebinarized) outputs into 16-bit packets.
\end{itemize}

In Fig.~\ref{fig:chewbaccaNNCore}, the compute core of ChewBaccaNN is depicted. It can be seen that several of the parameters listed in Fig.~\ref{fig:conv_params} are hard-wired into the design. The kernel height ($R$) and width ($S$) are completely unrolled (in this case with a factor of 7), while the channel dimension ($C$) should be a multiple of 16 (the number of \texttt{XNOR} gates) to achieve full utilization; in other words, the vectorization factors are $v_C = 16$, $v_R = 7$ and $v_S = 7$.

The Controlled Shift Register (CSR) allows to use a sliding window in order to get data reuse; for each IFM image row, initially, the full kernel width (in this case 7) is transferred, while the iterations thereafter only need one new column ($v_R \times 1 \times v_C$) of activations.

\subsubsection{Completely Unrolled Ternary Inference Engine (CUTIE)}

\begin{figure}[htb]
    \centering
    \includegraphics[width=0.9\linewidth]{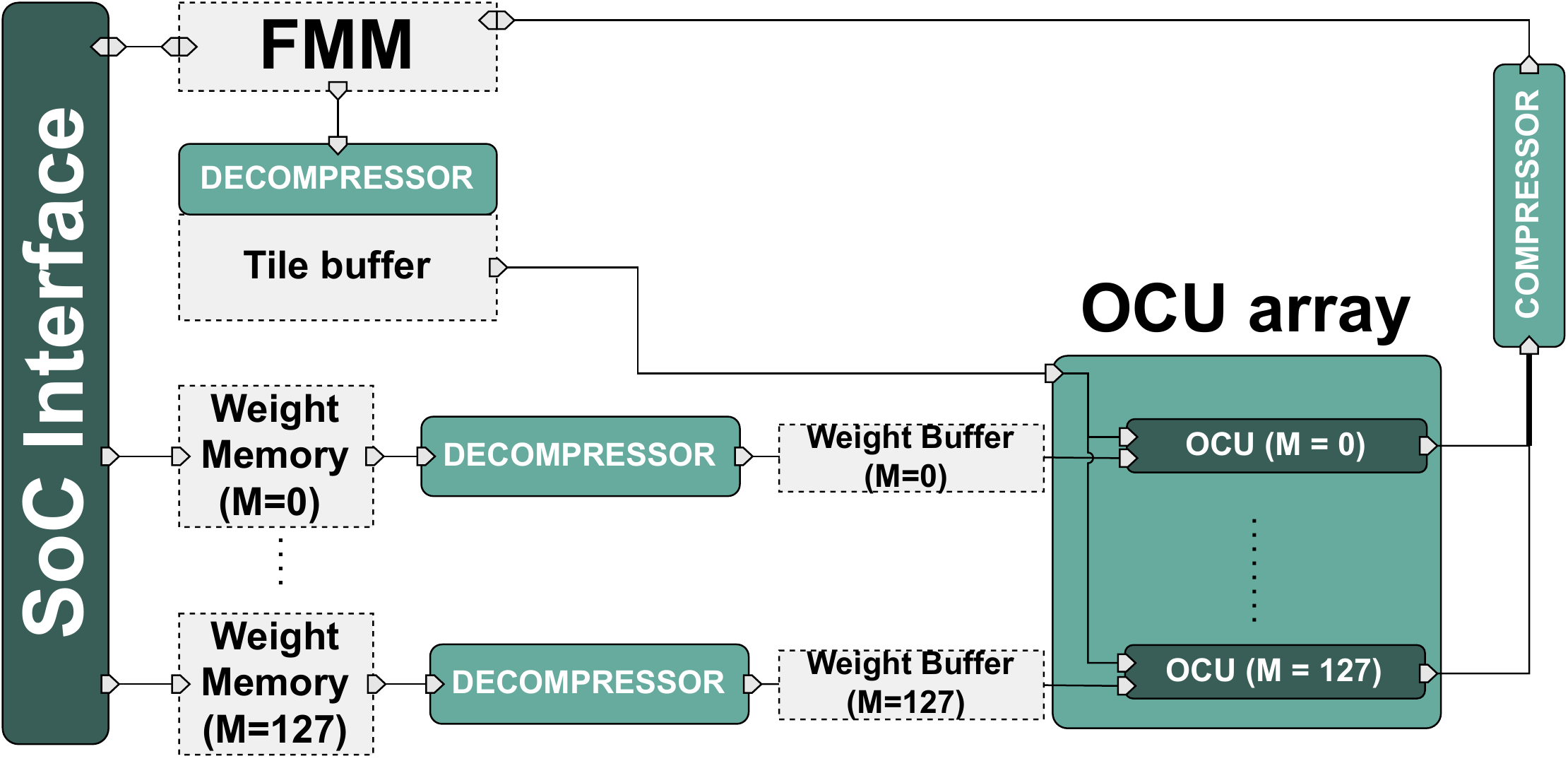}
    \caption{CUTIE top-level architecture. The OCU array contains one output channel compute unit for each output channel in the neural network design.}
    \label{fig:CUTIE}
\end{figure}

\begin{figure}[b]
    \centering
    \includegraphics[width=0.9\linewidth]{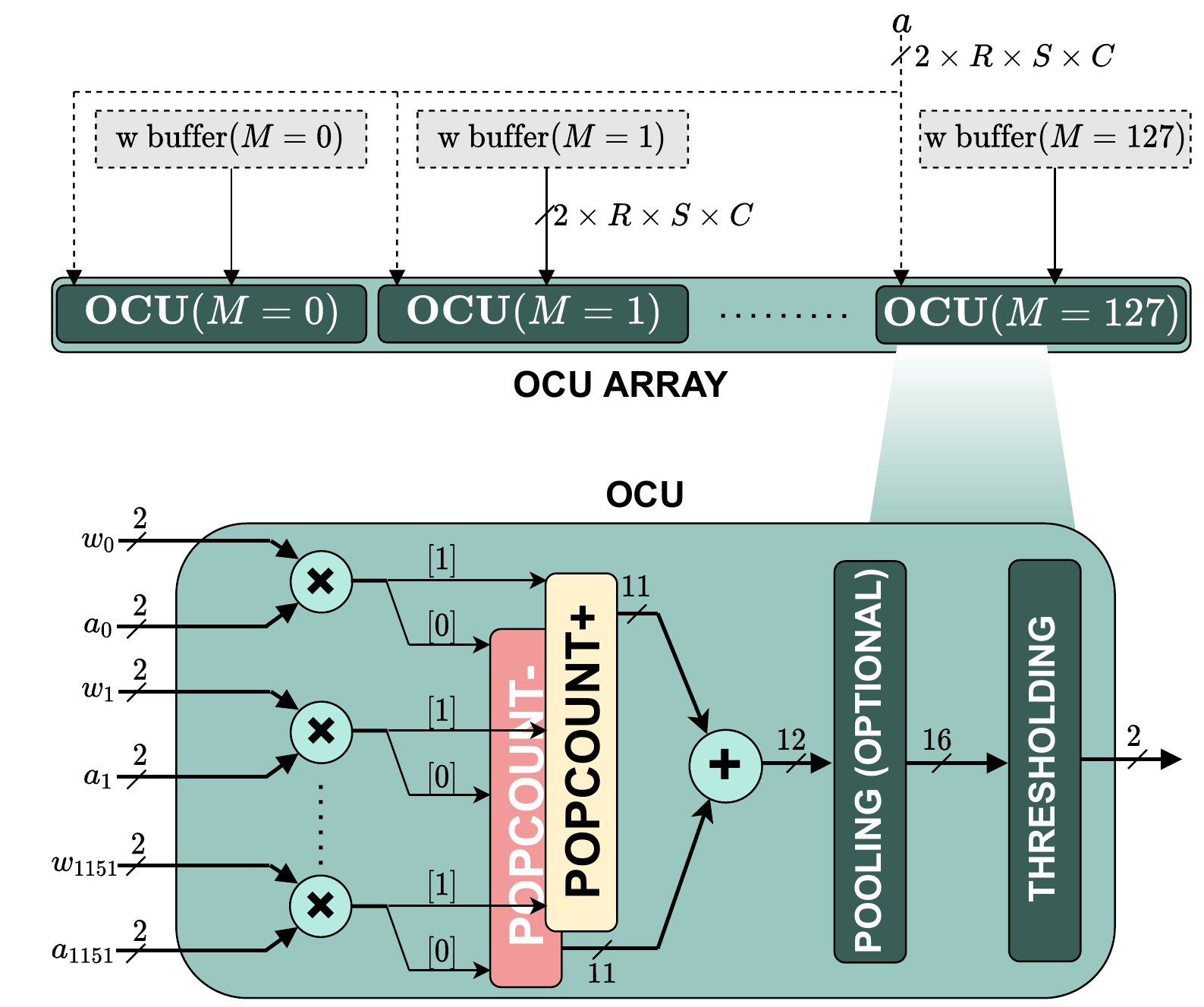}
    \caption{CUTIE compute core, consisting of several Output Channel Compute Units (OCUs) and one weight buffer per OCU. For brevity, decompression \& pipelining are omitted in this figure. The ternary multipliers are unrolled over the $R$, $S$ and $C$ dimension, which in this case gives $3 \times 3 \times 128 = 1152$ ternary multipliers. In total, the architecture can process $3 \times 3 \times 128 \times 128 = 147,456$ ($v_R \times v_S \times v_M \times v_C$) inputs each compute cycle.}
    \label{fig:CUTIEcore}
\end{figure}

Completely Unrolled Ternary Inference Engine (CUTIE)~\cite{Scherer2020CUTIE:Efficiency}, is, as the name suggests, an inference accelerator using \texttt{Ternary} operands. The main design philosophy behind CUTIE is to avoid iteration by spatially unrolling most of the convolutional loops found in Listing~\ref{lst:naive_cnn}, namely the loops over the $R$, $S$, $C$, and $M$ dimensions. Furthermore, \texttt{ternary} operands allow the representation of zero, therefore making the accelerator capable of exploiting neural network sparsity by silencing compute units. The top-level design of CUTIE is depicted in Fig.~\ref{fig:CUTIE}. The main components within the CUTIE architecture are:

\begin{itemize}
    \item \textit{Output Channel Compute Unit (OCU)}, the basic compute building block of this architecture, computing the output pixels belonging to one single output channel. A detailed view of the OCU is given in Fig.~\ref{fig:CUTIEcore}.
    \item \textit{Feature Map Memory (FMM)}, used to store the inputs coming either from previous computations (OCUs) or from an external interface. The FMM is double-buffered such that the latency for loading new input feature maps can be hidden.
    \item \textit{Tile buffer}, used to buffer IFM pixels in a sliding window fashion.
    \item \textit{Weight buffer}, one is attached to each OCU: it is designed with enough capacity to contain the full kernel for a single output channel ($R \times S \times C)$ which enables great weight reuse. The weight buffer is also double-buffered to hide latency.
    \item \textit{Compression/decompression units}, are used to shift between the computational form of the trits, i.e. 2-bits, and the compressed form of the trits which is 1.6-bits wide.
\end{itemize}

The compute core of CUTIE is depicted in Fig.~\ref{fig:CUTIEcore}. Its main workhorse is the Output Channel Compute Unit~(OCU), which is a unit that calculates pixels \textit{exclusive to a single output channel}. Having a separate compute unit for each output channel brings the advantage that the weight kernel can stay inside the weight buffer (w buffer) while moving the convolutional window over the IFM giving maximum weight data reuse. Alongside the weight reuse, there is also IFM reuse being utilized in two different ways: 1) the IFM is broadcasted to each of the OCUs, 2) just like ChewBaccaNN, when sliding the convolutional window over the IFM image, only $R$ new IFM pixels are needed (i.e. only one new column of the IFM needs to be loaded, assuming a stride of 1). 

Each Output Channel Compute Unit (OCU) processes $128 \times 3 \times 3$ ($v_C \times v_R \times v_S$) input pixels each cycle. By hard-wiring many of the convolutional layer parameters, CUTIE sacrifices area in favor of avoiding temporal iteration. This also means that this architecture sacrifices most flexibility by constraining $C$, $M$, $R$, and $S$. Therefore, the only dimensions that are freely schedulable are $W$ and $H$. By constraining many of the dimensions into the hardware, flexibility crumbles but the temporal mapping is greatly simplified, while the spatial mapping provides inherently great data reuse by design. Since each OCU directly computes an output pixel, there is no need, in contrast to the other architectures, to move around partial results. This is beneficial since the partial results always have a higher bit-width than the final (requantized) results.

\subsubsection{Binary Neural Network Accelerator in 10nm FinFet}

\begin{figure}[htb]
    \centering
    \includegraphics[width=0.9\linewidth]{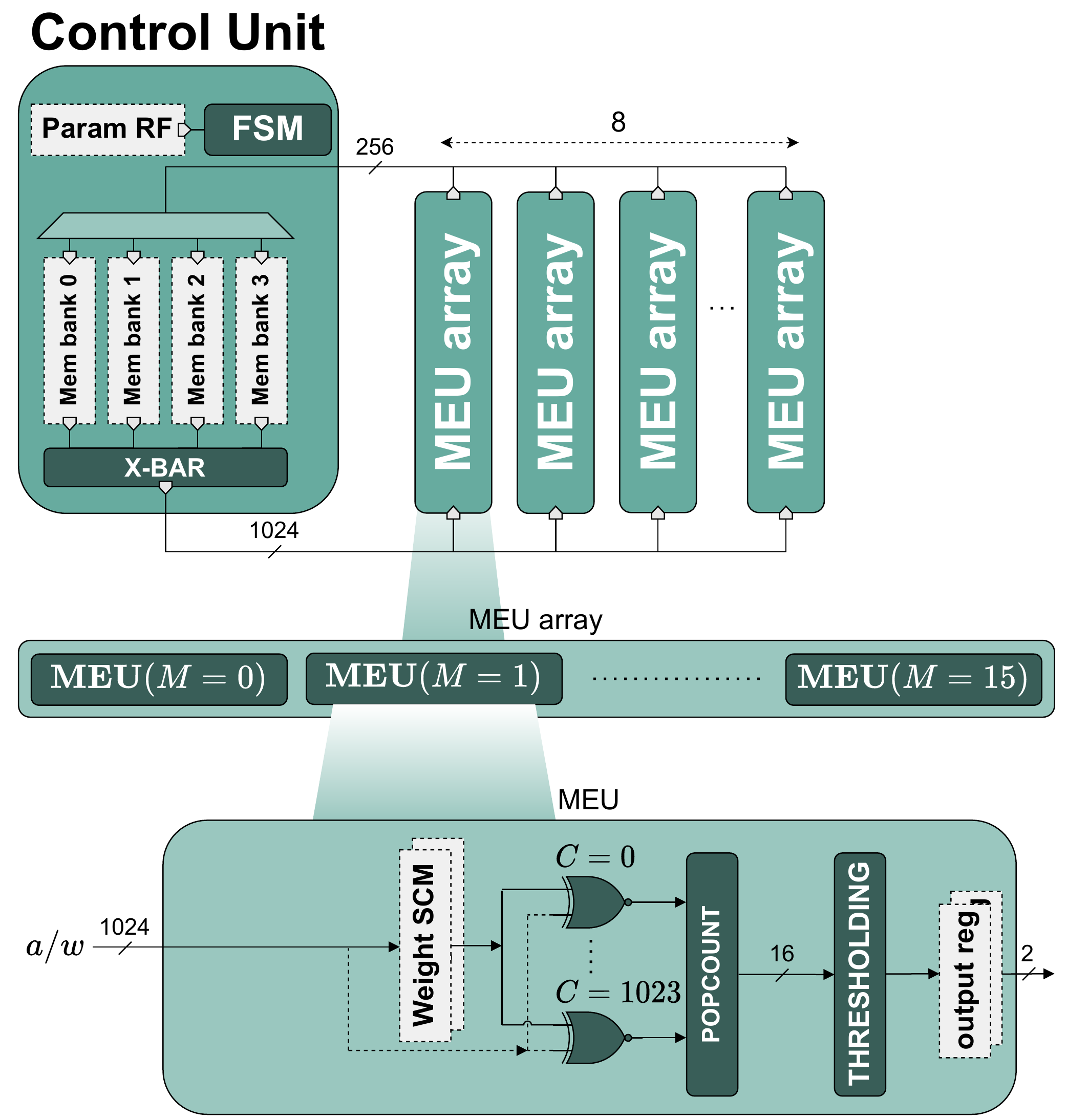}
    \caption{Top-level view of the 10nm FinFet BNN accelerator. The central memory inside the control unit consists of 4x 256-bit wide SRAM banks to enable 2x2 convolutional window access in a single cycle and a finite-state machine (FSM). The MEUs are placed in an 8x16 array to exploit the compute near memory principle. In total, $1024 \times 16 \times 8 = 131,072$ ($v_C \times v_M \times 8$) binary operations can be performed each cycle.}
    \label{fig:10nm_accelerator}
\end{figure}

In \cite{Knag2021ACMOS}, Knag et al. show a fully-digital accelerator with \texttt{binary} operands which is implemented using 10nm FinFet technology. The SoC designed intersperses arithmetic with memory according to the Compute Near Memory~(CNM) paradigm. Contrary to the other architectures discussed, this paper focuses more on the physical implementation and circuit-level design choices than the architectural design aspects. The design of this accelerator is shown in Fig.~\ref{fig:10nm_accelerator}. The main components of this accelerator are:
\begin{itemize}
    \item \textit{Control Unit}, consists of four 256-bit wide SRAM memory banks used as main storage and a Finite-State Machine (FSM) that controls the flow of data between memory banks and the MEUs. 
    \item \textit{Memory Execution Unit (MEU)}. Each MEU can compute two output pixels in a time-interleaved manner (see Fig.~\ref{fig:10nm_accelerator}, each MEU contains two output registers). The MEUs are interleaved with latch-based memories to utilize the compute near memory advantages. In total there is an array of $16 \times 8$ MEUs. Having 8 weight SCM banks was found to be the right trade-off between energy consumed by the computational elements and energy consumed by the transportation of data to the compute units. The SRAM memory banks are connected to the MEUs by means of a crossbar network. Since the input feature map pixels are stored in an interleaved manner, the crossbar network allows any (2x2) combination of the input feature map to be read. The weights are also loaded from this memory.
\end{itemize}

The authors of the paper do not discuss the external interfacing required on this chip. 

Since binary arithmetic is relatively cheap, to amortize the costs of memory reads and data movement, the computation should be scaled up to balance the energy consumption. When performing \texttt{binary MAC} operations, the results, which have a much higher bit-width than the inputs, need to be stored in an accumulator. The energy cost of this accumulator is very large with respect to the \texttt{xnor} gate. Therefore, to mitigate this cost, parallelism should be applied inside the \texttt{MAC} unit. Like the other architectures, this accelerator parallelizes the \texttt{MAC} operation over the input channel ($C$) dimension. The parallelization should be high enough to offset the accumulator cost while being low enough to not impose unreasonable constraints on the number of input channels ($C$) required for full utilization. Therefore, a trade-off study was performed to see which level of parallelism was needed to offset the accumulator cost. A design with an input feature map parallelization factor of 1024 ($v_C=1024$) was chosen as the sweet-spot. Negligible energy improvements were shown when going for more parallelism.

Furthermore, the idea of pipelining the \texttt{popcount-adder} tree was explored. When pipelining the design, the voltage can be lowered at iso-performance (i.e. iso-frequency). However, due to the sequential logic and clock-power dissipated while adding more pipeline stages, the final design choice was to not pipeline the \texttt{popcount-adder} tree.

\subsubsection{BrainTTA}

BrainTTA is a fully compiler-programmable mixed-precision flexible-datapath architecture. Contrary to the fixed-path accelerators, BrainTTA is based on the Transport-Triggered Architecture~(TTA)~\cite{Corporaal1997MicroprocessorTTA} that provides a fully programmable datapath (via a compiler) directly to the user. Before diving into the BrainTTA architecture, a proper introduction to the Transport-Triggered Architecture is given.

Transport-Triggered Architectures are programmed by data movements instead of arithmetic operations typically found in VLIW architectures. This means that the movement of data between function units (FUs) and register files (RFs) is exposed to the programmer; the TTA is an explicit-datapath architecture. This is in stark contrast to VLIW architectures, where the data movement is implicit and performed in hardware (i.e. not exposed to the programmer). With the control of the datapath given to the compiler, several optimizations can be performed like operand sharing and register file bypass.

\begin{figure}[ht]
    \centering
    \includegraphics[width=0.7\linewidth]{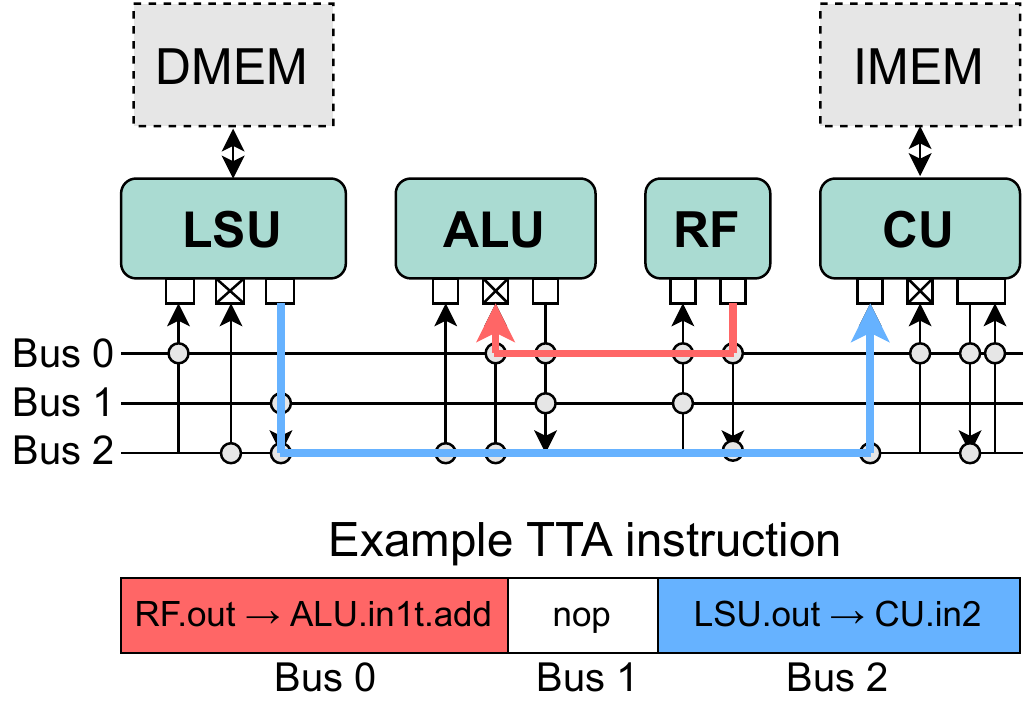}
    \caption{An example TTA instance and instruction, the square blocks denote \textit{input-} and \textit{output-ports}. A cross denotes a \textit{trigger-port}. The colored arrows drawn on the architecture illustrate the \textit{move operations} inside the example instruction.}
    \label{fig:ttaBasic}
\end{figure}

An example instance of a TTA is displayed in Fig.~\ref{fig:ttaBasic}. The TTA consists of a Control Unit (CU) used for instruction fetching and decoding, Register Files (RFs) for temporary storage, and Load-Store Units (LSUs) to access the memories. The grey circles inside the busses denote that this bus is connected to the corresponding input- or output-port of some function unit. This connectivity is design-time configurable, visible to the compiler, and can be made as generic or specific for certain applications as desired; more connectivity is at the expense of larger instruction size and more switching activity in the interconnect. In~\cite{Multanen2021Energy-EfficientProcessors}, Multanen presented several ways to alleviate this effect by applying techniques that reduce the instruction overhead such as instruction compression. An example instruction is shown in Fig.~\ref{fig:ttaBasic} which shows that the instruction can be broken down into \textit{move operations} for each bus.
\begin{figure}[ht]
    \centering
    \includegraphics[width=1.0\linewidth]{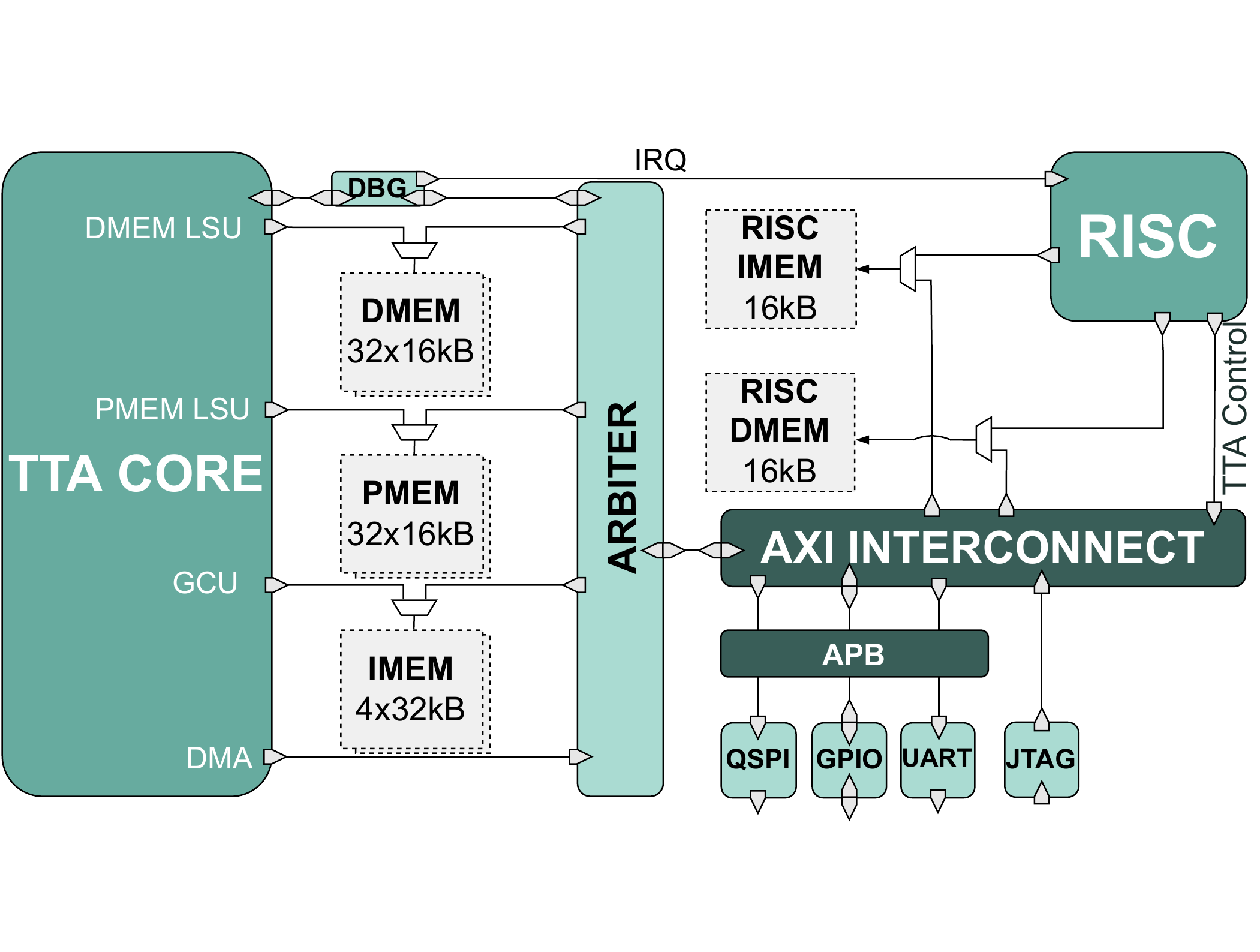}
    \caption{Top-level view of the BrainTTA SoC, the arbiter forms the border between the RISC and TTA part of the SoC.}
    \label{fig:braintta_block}
\end{figure}

BrainTTA is based on the TTA, built specifically for inference with precisions \texttt{integer8}, \texttt{binary} and \texttt{ternary}. A top-level view of the BrainTTA SoC is shown in Fig.~\ref{fig:braintta_block}. BrainTTA is designed using the open-source toolchain TTA-based Co-design Environment (TCE)~\cite{Esko2010CustomizedSupport}\cite{Jaaskelainen2016HW/SWProcessors}. The SoC consists of:

\begin{itemize}
    \item \textit{RISC-V host processor}, which is taken from an open-source repository~\cite{Traber2017PULPino:Datasheet}, the host processor starts and halts execution of the TTA core and takes care of the external communication (e.g. loading the on-chip memories).
    \item \textit{TTA core}, the workhorse of the architecture, supports mixed-precision inference.
    \item \textit{SRAM Memories}, separate memories for the RISC and TTA core, the TTA core memories are highly banked to allow efficient access of smaller bit-widths, while also supporting wide vector accesses. The TTA core is connected to three memories, the DMEM, used for storing input and output feature maps, the PMEM, used to store the weights, and the IMEM used for instructions to program the behavior of the TTA core.
    \item \textit{Debugger (DBG)}, used to control the execution of the TTA core, can signal task completion to the RISC-V.
    \item \textit{AXI interconnect}, used for on- and off-chip communication between the RISC, TTA-core, and peripherals. 
\end{itemize}

The workhorse of this architecture is the TTA core, where the actual inference happens. Details of the TTA core instantiation used in BrainTTA can be found in Fig.~\ref{fig:braintta_core}. The core contains different Function Units (FUs), divided into scalar and vector FUs. The FUs are interconnected via the busses, with 32-bit scalar busses (bus 0-5) and 1024-bit vector busses (bus 6-9). The core consists of the following units:

\textbf{Control Unit (CU)}; it contains the logic to fetch and decode instructions and steers the other units to execute the correct operations. Furthermore, the CU contains a hardware loopbuffer to save energy on the instruction memory accesses. This can be very beneficial since all network layers are essentially described by multiple nested loops (see Listing~\ref{lst:naive_cnn}).

\textbf{Vector Multiply-Accumulate (vMAC)}, the actual number cruncher. This unit supports the following operations: \texttt{integer8 MAC} (scalar-vector and vector-vector), \texttt{binary MAC} and \texttt{ternary MAC}. Its vector size is 1024-bit, with 32 entries of 32-bits each. The scalar-vector \texttt{MAC} multiplies a scalar by a vector by broadcasting the (32-bit) scalar value to all slots. This is beneficial when multiple inputs share the same weights (as in convolution).

For each precision \texttt{MAC} operation, the vectorization factor is different. All arithmetic circuitry contains 32 accumulators for the (intermediate) output channel result, i.e. $v_M = 32$. The number of concurrent input channels than is the vector size~(1024) divided by 32 (number of output channels) divided by the operand size (i.e. 1,2 or 8 bits). Therefore, the input channel vectorization is $v_C=32$, $v_C=16$ and $v_C=4$ for \texttt{binary}, \texttt{ternary} and \texttt{integer8} respectively.

\textbf{Vector Add (vADD)} is used to add two (either 512- or 1024-bit) vectors. This can e.g. be used to support residual layers.

\textbf{Vector Operations (vOPS)}, auxiliary (vector) operations that are required in the network, alongside to the computations. This FU can perform \texttt{requantization}, \texttt{binarization}, \texttt{ternarization}, as well as activation functions e.g. \texttt{ReLU} and pooling functions such as \texttt{MaxPool}. Furthermore, various other operations to extract and insert scalar elements into a vector are also supported by this unit.

\textbf{Register Files (RFs)} come in different bit-widths, namely binary, 32-bit scalar, and 1024-bit vector. These registers can be used to facilitate data re-use and store intermediate results without performing (more costly) access to the SRAM.

\textbf{Load-Store Units (LSUs)} form the interface between the TTA core and the SRAM memory. For each memory, there is a separate LSU to facilitate concurrent weight and input loading. The units support loads and stores for different bit-widths ranging from 8-bits all the way up to 1024-bits. Since the memory is banked, a strobe signal can be used to selectively turn on banks when data with smaller bit-widths are loaded/stored, in order to save energy.

\textbf{Scalar ALUs} are mostly used for address calculations needed as inputs to the LSUs. These units support basic arithmetic on values up to 32-bit.
\begin{landscape}
    \centering
    \vspace*{\fill}
    \begin{figure}[bh]
        \centering
        \includegraphics[width=1\linewidth]{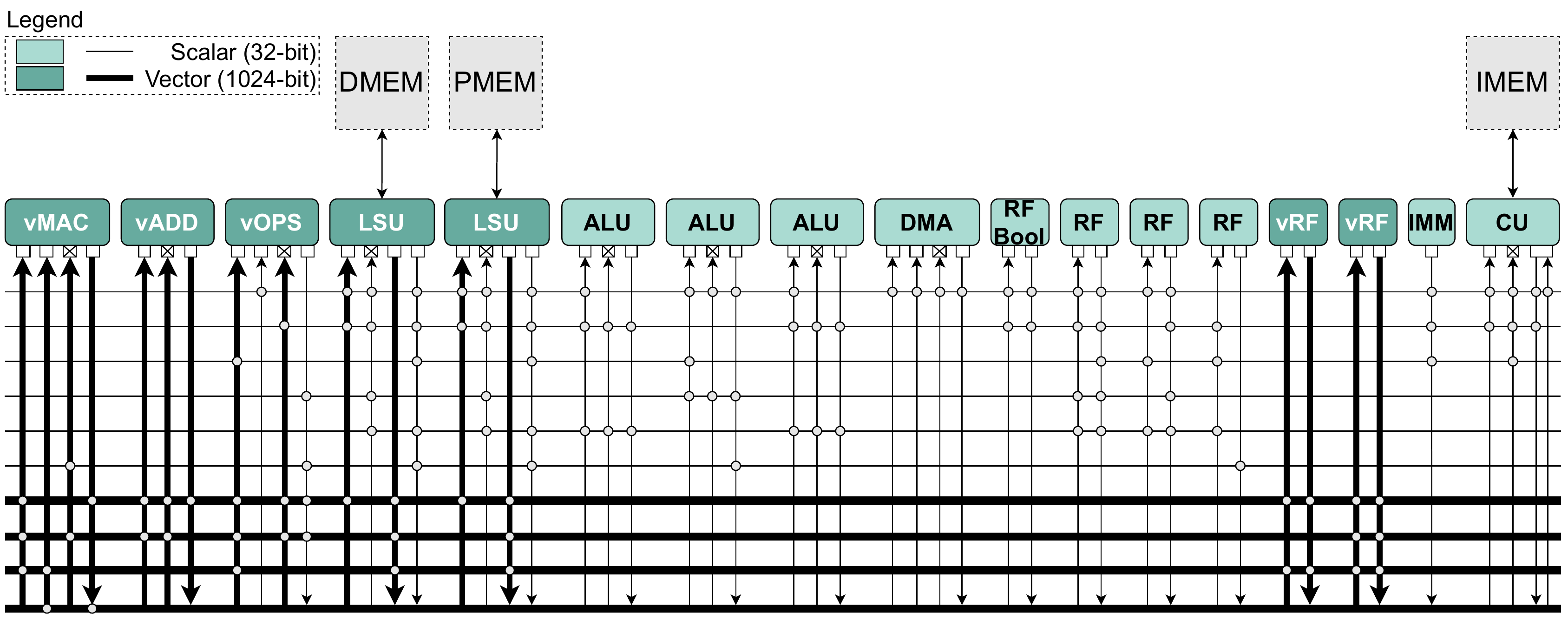}
        \caption{BrainTTA core instance, thicker lines denote vector busses, thinner lines scalar busses.}
        \label{fig:braintta_core}
    \end{figure}
    \vfill
\end{landscape}

\begin{table}[h]
\caption{Comparison of performance, efficiency and flexibility of the architectures discussed.}
\label{tab:sota_accelerators}
\centering
\begin{threeparttable}
\begin{tabular}{lrclcrrr}
 & \multicolumn{1}{c}{\begin{tabular}[c]{@{}c@{}}Chew-\\ BaccaNN~\cite{Andri2020ChewBaccaNN:Accelerator}\end{tabular}} & \multicolumn{2}{c}{CUTIE~\cite{Scherer2020CUTIE:Efficiency}} & \multicolumn{2}{c}{XNE~\cite{Conti2018XNORInference}} & \multicolumn{1}{c}{\begin{tabular}[c]{@{}c@{}}10nm\\ FinFet~\cite{Knag2021ACMOS}\end{tabular}} & \multicolumn{1}{c}{BrainTTA} \\ \hline \hline
\textbf{\begin{tabular}[c]{@{}l@{}}Implementation\\ characteristics\end{tabular}} & \multicolumn{1}{l}{} & \multicolumn{1}{l}{} &  & \multicolumn{1}{l}{} & \multicolumn{1}{l}{} & \multicolumn{1}{l}{} & \multicolumn{1}{l}{} \\
Tech node [nm] & 22 & \multicolumn{2}{c}{22} & \multicolumn{2}{c}{22} & 10 & 28 \\
Supply voltage [V] & 0.4 & \multicolumn{2}{c}{0.65} & \multicolumn{1}{r}{0.6} & 0.4 & 0.39 & 0.9 \\
Inference precision\tnote{1} & b & \multicolumn{2}{c}{b\tnote{2}, t} & \multicolumn{2}{c}{b} & b & b, t, i8 \\
Memory technology & SCM & \multicolumn{1}{r}{SRAM} & \multicolumn{1}{r}{SCM} & \multicolumn{1}{r}{SRAM} & SCM & SRAM & SRAM \\ \hline
\textbf{KPIs} & \multicolumn{1}{l}{} & \multicolumn{1}{l}{} &  & \multicolumn{1}{l}{} & \multicolumn{1}{l}{} & \multicolumn{1}{l}{} & \multicolumn{1}{l}{} \\
Peak throughput [GOPS] & 240 & \multicolumn{2}{c}{16000} & \multicolumn{1}{r}{67} & 5 & 3400 & 880 \\
Energy/op [fJ] binary & 4.48/15.38\tnote{3} & \multicolumn{2}{c}{-} & \multicolumn{1}{r}{115} & 21.6 & 1.62 & 101 \\
Energy/op [fJ] ternary & - & \multicolumn{1}{r}{2.19} & \multicolumn{1}{r}{1.70} & \multicolumn{2}{c}{-} & - & 188 \\
Energy/op [fJ] 8-bit & - & \multicolumn{2}{c}{-} & \multicolumn{2}{c}{-} & - & 1105 \\
Core area [mm$^2$] & 0.7 & \multicolumn{2}{c}{7.5} & \multicolumn{2}{c}{2.32} & 0.39 & 3.6 \\
Area efficiency [GOPS/mm$^2$] & 343 & \multicolumn{2}{c}{2133} & \multicolumn{2}{c}{28.88} & 8717 & 244.44 \\
Memory capacity [kB] & 153 & \multicolumn{1}{r}{1190} & \multicolumn{1}{r}{281} & \multicolumn{1}{r}{520} & 16 & 161 & 1024 \\ \hline
\textbf{Flexibility} &  & \multicolumn{1}{r}{} &  & \multicolumn{1}{r}{} & \multicolumn{1}{l}{} &  &  \\
Full utilization for & \multicolumn{1}{l}{} & \multicolumn{1}{l}{} &  & \multicolumn{1}{l}{} & \multicolumn{1}{l}{} & \multicolumn{1}{l}{} & \multicolumn{1}{l}{} \\
\quad C multiple of & 16 & \multicolumn{2}{c}{128} & \multicolumn{2}{c}{128} & 1024 & 32/16/4\tnote{4} \\
\quad M multiple of & Any & \multicolumn{2}{c}{128} & \multicolumn{2}{c}{128} & 128 & 32 \\
\quad R is & 7 & \multicolumn{2}{c}{3} & \multicolumn{2}{c}{Any} & 2 & Any \\
\quad S is & 7 & \multicolumn{2}{c}{3} & \multicolumn{2}{c}{Any} & 2 & Any \\
\begin{tabular}[c]{@{}l@{}}Partial result support\\ (for scheduling freedom)\end{tabular} & Yes & \multicolumn{2}{c}{No\tnote{5}} & \multicolumn{2}{c}{No} & No & Yes \\
Residual layer support & Yes & \multicolumn{2}{c}{No} & \multicolumn{2}{c}{No} & No & Yes \\
Programmability & None & \multicolumn{2}{c}{None} & \multicolumn{2}{c}{None} & None & C-language \\ \hline
\end{tabular}
\begin{tablenotes}
    \item[1] b = \texttt{binary}, t = \texttt{ternary}, i8 = \texttt{integer8}.
    \item[2] Only estimations were provided, under the assumption that all ternary specific hardware is removed.
    \item[3] For 7x7 and 3x3 convolution respectively.
    \item[4] For \texttt{binary}, \texttt{ternary} and \texttt{integer8} respectively.
    \item[5] Partial result support is not needed since the output pixel computation is fully unrolled in hardware.
\end{tablenotes}
\end{threeparttable}
\end{table}
\section{Comparison \& Discussion}
\label{sec:comp}

All architectures discussed in Section~\ref{sec:accel} are evaluated on flexibility and energy efficiency. These results are given in Table~\ref{tab:sota_accelerators}. This table is split into three sections: the implementation characteristics, performance characteristics as discussed in Section~\ref{sec:kpi}, and the flexibility aspects as discussed in Section~\ref{sec:flex}.

The energy efficiency of the accelerators ranges from 1.6~fJ to 115~fJ per operation for binary precision, a large range. It should be noted, however, that the two architectures that have the highest energy usage (XNE and BrainTTA), are the only architectures that show a full autonomous SoC including peripherals. Furthermore, all architectures except BrainTTA utilize voltage-frequency scaling to run the accelerator at lower than nominal supply voltage, trading off throughput for better energy efficiency.

Next to the energy efficiency, the table also lists the neural network layer requirements that these architectures impose in order to fully utilize the arithmetic hardware. It is seen that the most energy-efficient architectures, CUTIE~\cite{Scherer2020CUTIE:Efficiency} and the BNN accelerator in 10nm FinFet from Knag et al.~\cite{Knag2021ACMOS} are also the most constrained architectures, in terms of neural network layer requirements. Therefore, the question arises, does hard-wiring the neural network layer parameters directly improve the energy efficiency of an architecture, for different models, also when layer variety is high? 

Interestingly, the XNE and BrainTTA share very similar layer constraints. Both are only constrained in the input channel ($C$) and output channel ($M$) dimensions. The energy consumption of BrainTTA is somewhat lower at an older technology node while using a higher supply voltage. The reason for this is that BrainTTA better exploits data reuse. The execution schedule for BrainTTA was tuned to maximize data reuse, while XNOR neural engine only reuses a set of input feature maps for \texttt{TP} (in this case 128) cycles while reloading the weights for each \texttt{MAC} operation.

XNE was also benchmarked using SCM only, severely cutting the very high energy cost associated with these redundant memory fetches, at the cost of losing memory capacity. Some architectures report energy numbers for an SCM as well as an SRAM implementation. The memory capacity of the SCM versions is very low compared to the SRAM versions, hindering the ability to run full-size networks on it without adding expensive off-chip memory accesses. For the sake of comparison, for all the architectures with an SRAM version available, the SRAM version is chosen for further analysis.

Support for residual layers can only be found in ChewBaccaNN and BrainTTA. Other architectures are not able to support this due to their fixed datapath. The data-flow through these accelerators is very static and the accumulated value will directly be \texttt{binarized} or \texttt{ternarized} after all inputs are accumulated. This prohibits the use of residual layers since residual layers need the intermediate (larger bit-width) results that were obtained before requantization.

It is clear that parallelism and data reuse (either in the form of locally buffering or by broadcasting) are the keys to amortizing the memory access cost, which is so much larger than the low-precision arithmetic cost. Techniques to mitigate these costs are to replace SRAM with low-voltage SCM, hard-wire network parameters to enable broadcasting, and use the sliding window principle (like the FMM banks in combination with the crossbar in ChewBaccaNN~\cite{Andri2020ChewBaccaNN:Accelerator}). In essence, all these solutions boil down to designing the architecture around the data movements in a less-flexible manner. These architectures solve the mapping problem by fixing most parameters using \textbf{spatial mapping}, greatly simplifying the task of \textbf{temporal mapping} at the cost of losing flexibility. XNE and BrainTTA fix the least number of parameters using \textbf{spatial mapping}, therefore leaving a larger \textbf{temporal mapping} space to be explored. 

\section{Summary \& Conclusions}
\label{sec:conc}

Neural networks are all around and are making an advance into the embedded domain. With the increasing popularity of edge computing, new methods are needed to port the typically power- and memory-hungry neural networks to devices that have limited storage and are subject to severe energy constraints. Quantization is a fundamental ingredient in overcoming these challenges. Very low precisions, down to 1-bit have shown to achieve great energy efficiency while drastically reducing the model size and computational cost involved in neural network inference. To fully exploit the reduced computational complexity and memory requirements of these networks, neural network accelerators aimed specifically at these heavily quantized networks have been developed.

In this chapter, state-of-the-art low- and mixed-precision architectures are reviewed. Taking into account the variety present in network layers of CNNs, the architectures are compared against each other in terms of flexibility and energy efficiency. It was found that spatially mapping more dimensions of the neural network layer increases the energy efficiency as it allows minimization of data movement by tailoring the memory hierarchy design, which is a big contributor to energy cost in inference accelerators. Contrary to the group of accelerators that maps most layer dimensions spatially, there is a group of accelerators that minimizes the layer dimension requirements by less heavily relying on spatial mapping, retaining more freedom in the temporal mapping domain. They are more flexible and can handle a larger part of the neural architecture design space. In addition, they may have support for multiple bit precisions.

With new attempts to streamline the process of finding the best combination of temporal- and spatial mappings~\cite{Mei2021ZigZag:Accelerators}, while co-designing the memory hierarchy, the question arises if an optimized temporal mapping in combination with memory hierarchy co-design can close the energy efficiency gap with the more constrained, heavily spatially mapped accelerators, giving better energy efficiency at a wider range of neural network layers.

\bibliographystyle{style/spmpsci}
\bibliography{references.bib}

\end{document}